\documentclass[%
 reprint,
 amsmath,amssymb,
 aps,
]{revtex4-2}

\usepackage{graphicx}
\usepackage{dcolumn}
\usepackage{bm}
\usepackage{dsfont}
\usepackage[hidelinks]{hyperref}
\usepackage[capitalise]{cleveref}
\usepackage{tabularx}
\usepackage[dvipsnames]{xcolor}
\usepackage{enumitem}
\usepackage{makecell}
\usepackage{array}

\begin{document}

\preprint{APS/123-QED}

\title{Information Benchmark for Biological Sensors Beyond Steady States -- Mpemba-like sensory withdrawal effect}

\author{Asawari Pagare}
 \affiliation{Department of Chemistry, University of North Carolina-Chapel Hill, NC}
\author{Zhiyue Lu}%
 \email{zhiyuelu@unc.edu}
\affiliation{%
Department of Chemistry, University of North Carolina-Chapel Hill, NC}%

\date{\today}

\begin{abstract}
Biological sensors rely on the temporal dynamics of ligand concentration for signaling. The sensory performance is bounded by the distinguishability between the sensory state transition dynamics under different environmental protocols. This work presents a comprehensive theory to characterize arbitrary transient sensory dynamics of biological sensors. Here the sensory performance is quantified by the Kullback-Leibler (KL) divergence between the probability distributions of the sensor's stochastic paths. We introduce a novel benchmark to assess a sensor's transient sensory performance arbitrarily far from equilibrium. We identify a counter-intuitive phenomenon in multi-state sensors: while an initial exposure to high ligand concentration may hinder a sensor's sensitivity towards a future concentration up-shift, certain sensors may show a boost in sensitivity if the initial high concentration exposure is followed by a transient resetting at a low concentration environment. The boosted performance exceeds that of a sensor starting from an initially low concentration environment. This effect, reminiscent of a drug withdrawal effect, can be explained by the Markovian dynamics of the multi-state sensor, similar to the Markovian Mpemba effect. Moreover, an exhaustive machine learning study of 4-state sensors reveals a tight connection between the sensor's performance and the structure of the Markovian graph of its states.

\end{abstract}

\maketitle


\section{\label{sec:inro-level1}Introduction}
Sensory receptors perceive information from external environments and transmit it into the cell via various signaling mechanisms, despite noise due to thermal fluctuations or imperfections \cite{antebi2017combinatorial, antebi2017operational, lestas2010fundamental, hinczewski2014cellular}. A ligand-receptor sensor that reports the level of ligand concentration is a classic example of biological sensor operating in the stochastic regime. The accuracy of ligand-receptor sensory mechanisms has been intensively studied in the steady-state regime \cite{berg1977physics, bialek2005physical, kaizu2014berg,singh2015accurate, nguyen2015receptor, bialek2008cooperativity, bialek1997statistical}.  However, recent studies of various biological processes -- e.g., extracellular signal-Regulated Kinase (ERK) pathway \cite{marshall1995specificity}, and NF-kB signalling under inflammatory stimuli \cite{hoffmann2002ikappab} --  have revealed that cells respond differently to different temporal patterns of external signals \cite{purvis2013encoding,mora2019physical, hopfield1974kinetic, behar2010understanding,  qian2006reducing}. Concepts from information science, such as mutual information, Shannon entropy, cross entropy, Kullback–Leibler (KL) divergence, etc., have been explored and utilized for analyzing the signaling capacity of various biological processes \cite{murugan2012speed, bauer2023information, bialek2012biophysics, potter2017dynamic, tang2021quantifying, cepeda2019estimating, hahn2023dynamical, pagare2023theoretical}. 

\begin{figure*}
\includegraphics{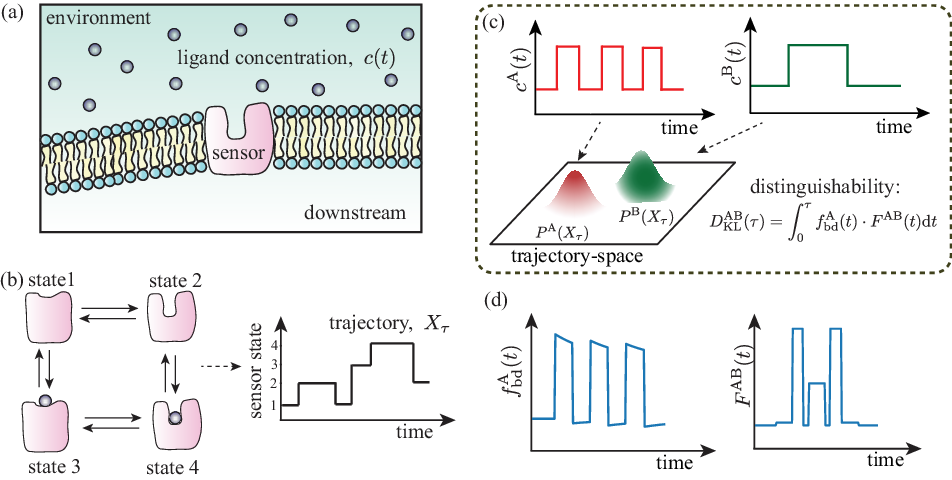}
\caption {(a) A ligand-receptor sensor on the surface of a cell probes the environment through binding with ligands and undergoes conformational changes that lead to downstream signalling, (b) a sensor can have multiple bound and unbound states and the temporal trajectory of the sensor's state contains richer information about the environment than in the state probability distribution \cite{tang2021quantifying, cepeda2019estimating, hahn2023dynamical, pagare2023theoretical}, (c) under different temporal patterns of the environmental ligand concentration different ensembles of stochastic trajectories can be observed. The distinguishability between the two ensembles is given by their KL divergence. (d) The KL divergence is shown to be equal to the time integral of the product between the binding frequency, $f^{\rm A}_{\rm bd}(t)$ weighted by the factor $F^{AB}(t)$ as shown in \cref{eq:KL_sensor}.}
\label{fig:sensor_trajectory}
\end{figure*}

The sensor's states and the transitions among those states may play an important role in information sensing, particularly temporal pattern recognition. As pointed out in references \cite{smith2018biased, gregorio2017single, che2020nanobody, zhang2023fast, thomas2023structure, han2023antigen, shen2019conformational, de2014allosteric, chen2017elucidation, hubbard2007receptor, volinsky2013complexity, sieghart2015allosteric, may2007allosteric}, even simple ligand-receptor sensors contains more states than bound and unbound states. The sensor's transitions among various configurations may transduce more information to the downstream information sensory network than that of a binary-state sensor. Furthermore, a larger number of states allow for more possible  binding and unbinding pathways of the receptor. For example, Deupi and Kobilka use energy landscape to argue that the sensor may take different pathways to bind with a ligand \cite{deupi2010energy}.  Beyond the spatial complexity of the sensor's state space, the sensor's transient dynamics richer in information compared to the steady states \cite{tang2021quantifying,cepeda2019estimating, hahn2023dynamical, pagare2023theoretical}

These works on the spatial and temporal complexity of sensor dynamics inspire questions around the performance limitations of sensors of different state space and the corresponding kinetics. For example, a multi-state sensor may not necessarily take the same state transition path in the binding process and an unbinding process (see \cref{fig:sensor_trajectory}). Thus, its sensory capability for an up-shift and a down-shift of ligand concentrations may be significantly different. A further question that arises is whether a sensor that is sensitive to concentration up-shift can also be sensitive to a down-shift? What are the underlying design principles to ensure sensitivity in both ways? Beyond the stationary regime, how does the transient dynamics of a sensor affect the sensory capability of a sensor? How quick can a sensor recover from a previous exposure to a high or low ligand concentration? 

This paper aims to provide a theoretical framework to address the questions listed above. In particular, how does the connectivity and transition rates between the various states of a sensor affect its sensory capability. This work defines a general formula for the transient sensory upper limit of an arbitrary sensor. When applied to ligand-receptor sensors, we propose benchmark protocols to reveal a sensor's sensory capability as well as the recovery capability. The recovery capability refers to the ability of a sensor to reset its sensory capability after a previous exposure to a high or low ligand concentration environment. 
Furthermore, this paper identifies a general type of anomalous sensory behavior of ligand-receptor sensors: while an initial exposure to high ligand concentration may hinder a sensor's sensitivity towards a future concentration up-shift, certain sensors may show a boost in sensitivity if the initial high concentration exposure is followed by a transient resetting at a low concentration environment. The boosted performance exceeds that of a sensor starting from an initially low concentration environment. We name this type of behavior a {\it sensory withdrawal effect}. Finally, we employ machine learning to classify the sensory state graphs based on their structural features, revealing a strong correlation between the sensor's performance, its ability to exhibit the sensory withdrawal effect, and the structure of its Markovian state graph.

\section{Theoretical Framework}
\label{sec:II}
\subsection{Stochastic description of sensors}
We utilize Markov state model to capture the state-transition dynamics of an arbitrary sensor. For illustration, a 4-state sensor with two bound states and two unbound states is shown by the graph in \cref{fig:sensor_trajectory}a,b. For any ligand-receptor sensor, we can classify the states into unbound states and various $k$-bound states. Here $k$-bound states include singly bound state ($k=1$), doubly bound state ($k=2$), and so on. We assume that the transitions among the unbound states and the transitions among the bound states of same $k$ value all exhibit rates that are independent of the ligand concentration. The binding transitions, from an unbound state to a $(k=1)$-bound state, or from a $k$-bound state to a $(k+1)$-bound state, can be expressed by the product between the ligand concentration and the corresponding transition rate constant. In this paper, we denote the transition rate from state $j$ to state $i$ by $R_{ij}$. 

The sensor, given the external signal of ligand concentration $c(t)$, evolves according to time dependent transition rates $\{R_{ij}(t)\}$. The stochastic dynamics of a sensor results in state-transition trajectories $X_\tau$ that follows the probability distribution $P[X_\tau]$. Here $\tau$ denotes the length of the trajectory, which can be considered as the observation time. The ability for the downstream signalling pathway to distinguish temporal patterns of two external signals $c^{\rm A}(t)$ and $c^{\rm B}(t)$ via information from the sensor accumulated in the time period $\tau$ is thus limited by the distinguishability of the trajectory probabilities, see \cref{fig:sensor_trajectory}c. 

\subsection{Transient sensory limit}
We propose to use the trajectory Kullback–Leibler (KL) divergence as a universal characterization of a biological sensor's ability to distinguish different temporal patterns of external signal. By definition, this quantity characterizes the difference between the probability distributions of the sensors' transition pathways under two temporal protocols (two time-dependent signals ${\rm A}$ and ${\rm B}$):
\begin{equation}
    D^{\rm AB}(\tau) \equiv \int {\mathcal D}_ {X_{\tau}} P^{A}[X_{\tau}] \ln \frac{P^{A}[X_{\tau}]}{P^{B}[X_{\tau}]}
    \label{eq:DKL_general}
\end{equation}
where $P^{A}[X_{\tau}]$ denotes the probability for a sensor to undergo transition path $X_\tau$ within the observation duration $\tau$ (see \cref{fig:sensor_trajectory}.c), and the path integral takes all possible stochastic trajectories of sensor's state into consideration. This quantity serves as an upper limit of the temporal pattern distinguishability information that a sensor could pass to the downstream sensory networks. 

According to a recent theory  \cite{pagare2024stochastic}, for any transient process that is arbitrarily far from the steady state, the trajectory KL divergence can be expressed as an accumulated weighted sum of all observed transition events:
\begin{equation}
    D_{{\rm KL}}^{\rm AB}(\tau) = \sum_{\langle x, x'\rangle} \int_0^ \tau    J^{\rm A}_{x'x}(t) \cdot F^{\rm AB}_{x'x}(t)~{\rm d} t
    \label{eq:simu}
\end{equation}
where $J^{\rm A}_{x'x}(t)=R^{A}_{x'x}(t) \cdot p^{\rm A}(x;t)$ is the detailed probability current for sensor's transition from state $x$ to $x'$ at time $t$ under signal ${\rm A}$; the weighting factor $F_{x'_k x_k}^{\rm AB}(t)$ characterizes the transition rate difference for each transition at time $t$ for the sensor's dynamics under the two signals (see \cref{fig:sensor_trajectory}.d). In this master equation description, the probability of the sensor being in state $x$ at time $t$ under the protocol ${\rm A}$ is denoted by $p^{\rm A}(x;t)$.

For a ligand receptor sensor under two different signals, $c^{\rm A}(t)$ and $c^{\rm B}(t)$, their trajectory KL divergence can be significantly simplified into (see \cref{sec:level1}): 
\begin{equation}
    D^{ \rm AB}_{\rm KL}(\tau) = \int _0^\tau  f^{\rm A}_{\rm bd}(t) \cdot F^{\rm AB}(t) {\rm d} t
    \label{eq:KL_sensor}
\end{equation}
where $f^{\rm A}_{\rm bd}(t)$ is simply the total ligand binding frequency at time $t$ for the sensor under signal ${\rm A}$, and the weighting factor $F^{\rm AB}(t)$ is a simple function of the ligand concentrations under the two signals at time $t$:
\begin{equation}
    F^{\rm AB}(t) =  \ln {\frac{c^{\rm A}(t)}{c^{\rm B}(t)}} + \frac{c^{\rm B}(t)}{c^{\rm A}(t)} - 1
    \label{eq:weighting_factor}
\end{equation}
This result implies that, for an arbitrary sensor, the more binding events that likely occur during the time of a large weighting factor $F^{\rm AB}(t)$ (i.e., large signal difference), the more the sensor can distinguish the two temporal patterns of signals ${\rm A}$ and ${\rm B}$.

In summary, the trajectory KL divergence, especially \cref{eq:KL_sensor}, provides a convenient and intuitive way to understand the transient sensory capacity of a sensor to discern different temporal patterns of external signal. It leads to intuitive design rules to enhance the distinguishability: the better sensors are those with higher binding event frequencies when the ligand concentration difference between the two protocols are prominent. Moreover, this formula allows us to characterize the transient response of a sensor to arbitrary temporal signals of ligand concentration, and to study the sensor's transient response speed and recovery speed when it experiences sudden changes of ligand concentration. 

Note that \cref{eq:DKL_general} and \cref{eq:simu} apply to any sensor that senses arbitrary physical quantity, whereas \cref{eq:KL_sensor} and \cref{eq:weighting_factor} apply to any ligand-receptor sensors. In the next section we develop a benchmark protocol that can be used to study the performance of sensors in sensing a sudden concentration shift.

\section{Transient Sensory Responses and Benchmark}

\subsection{Sensory Response: Stationary Versus Transient}
Firsly, the proposed transient theory can be reduced to describe sensors at steady states. Let us start by considering the sensor's stationary difference under constant high and constant low ligand concentrations, $c_h$ and $c_l$. For two identical copies of a sensor, one at the stationary state of high ligand concentration $c_h$ and another at the stationary state of low ligand concentration $c_l$, they produces two different ensembles of state transition trajectories. The distinguishability between their stochastic trajectories is characterized by a symmetrized KL divergence modified from \cref{eq:KL_sensor}:
\begin{equation}
    D_{\rm ss}(\tau) = D^{\rm AB}_{\rm ss}(\tau)+D^{\rm BA}_{\rm ss}(\tau) = \tau \cdot (f^{\rm A}_{\rm ss}F^{\rm AB}+f^{\rm B}_{\rm ss}F^{\rm BA})
\end{equation}
where the $f^{\rm A}_{\rm ss}$ denotes the stationary binding event frequency at the stationary concentration $c_h$ and $F^{\rm AB}$ is defined by \cref{eq:weighting_factor} with $c^{\rm A}(t)=c_h$ and $c^{\rm B}(t)=c_l$. In other words, at the steady state, the sensor's ability to distinguish high and low ligand concentrations can be considered as the product between the observation time $\tau$ and a constant information accumulation rate 
\begin{equation}
    \dot D_{\rm ss}=f^{\rm A}_{\rm ss}F^{\rm AB}+f^{\rm B}_{\rm ss}F^{\rm BA}.
    \label{eq:d_dot_ss}
\end{equation}
This agrees with the intuition that the longer the observation time, the more one can distinguish the two environments. 

\begin{figure}
\includegraphics{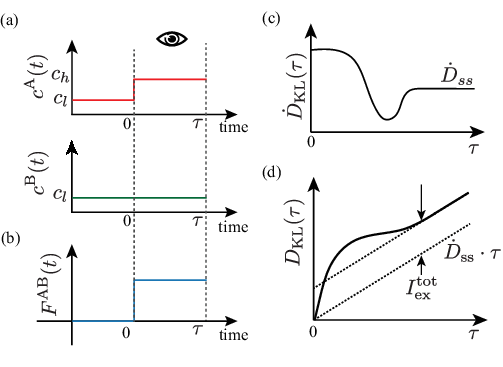}
\caption{(a) Protocols to characterize the distinguishability between a step-wise concentration up-shift (protocol A) from a constant concentration (protocol B). (b) The resulting weighting factor. (c) The rate of KL divergence as a function of time. Note that the rate is always positive but not necessarily monotonic; after long enough time from the initial concentration up-shift, it eventually settles to a constant rate $\dot D_{\rm ss}$. (d) The KL divergence as a function of observation time is monotonic with a positive information accumulation. The difference between the sensor's total accumulated information and the information accumulated under steady state information rate for the same time period defined the total excess information as in \cref{eq:i_tot_ex}.}
\label{fig:dkl_illustration}
\end{figure}

Then, in this work we go beyond the steady states and analyze the transient sensory capability of a sensor under the theoretical regime introduced in \cref{sec:II}. To simplify the discussion, we focus on a sensor's ability to distinguish a step-wise change of ligand concentration (see \cref{fig:dkl_illustration}a). In this case, consider two sensors initialized at the same initial stationary state at concentration $c_l$. Then at time $t=0$, one protocol introduces a sudden up-shift of concentration to $c_h$, whereas the other remains at $c_l$. The two sensors start to generate distinct trajectory probabilities after time $t=0$. By utilizing \cref{eq:KL_sensor}, we can find that the distinguishability between the two protocols, under observation period $(0,\tau)$ becomes a function of observation length $\tau$. This distinguishablibilty increases over $\tau$ with an {\it information accumulation rate}
\begin{equation}
    \dot D_{\rm KL}(\tau)= f^{\rm A}_{\rm bd}(\tau) F^{\rm AB} + f^{\rm B}_{\rm bd}(\tau) F^{\rm BA} 
\end{equation}
where $f^{\rm A}_{\rm bd}(\tau)$ and $f^{\rm B}_{\rm bd}(\tau)$ are the transient binding frequencies for the two protocols at time $\tau$ and the weighting factors $F^{\rm AB}$ and $F^{\rm BA}$ are both positive constants. \cref{fig:dkl_illustration}c illustrates the positive information accumulation rate as a function of observation length $\tau$. \cref{fig:dkl_illustration}d  illustrates the distinguishability $D_{\rm KL}(\tau)$, i.e., {\it accumulated information}, as a function of observation length $\tau$.

Given the above analysis, we can compare a sensor's transient sensory capability with the stationary sensory capability. As illustrated in \cref{fig:dkl_illustration}c, after a concentration up-shift, the transient response of a sensor may relax to the new steady state, and as a result the information accumulation rate converges to the steady-state information rate (\cref{eq:d_dot_ss}):
\begin{equation}
    \lim_{\tau\rightarrow \infty} \dot D_{\rm KL}(\tau) = \dot D_{\rm ss} \approx \dot D_{\rm KL}(\tau_{\rm ss})
\end{equation}
where $\tau_{\rm ss}$ denotes the relaxation time of the sensor. To highlight the difference between the sensor's transient sensory capability and steady-state sensory capability, we define their difference as the {\it total excess information}:
\begin{equation}
    I_{\rm ex}^{\rm tot} = \lim_{\tau\rightarrow \infty} D_{\rm KL}(\tau)- D_{\rm ss}(\tau)
    \label{eq:i_tot_ex}
\end{equation}
as shown in \cref{fig:dkl_illustration}d.
If $I_{\rm ex}^{\rm tot}>0$ the sensor's transient response to a sudden concentration up-shift is better than the steady-state sensory capability. If $I_{\rm ex}^{\rm tot}<0$ the sensor's transient response is worse than steady state.

\subsection{Ligand-receptor Sensor Benchmark}

\begin{figure}
\includegraphics{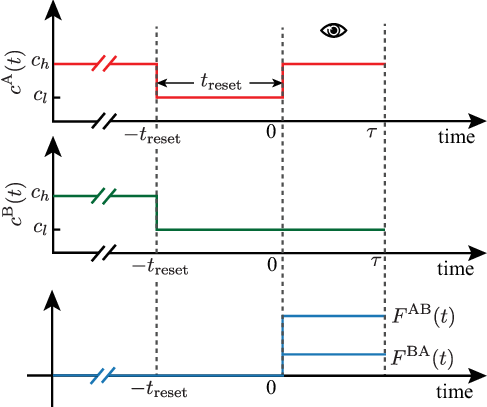}
\caption {Benchmark protocols $c^{\rm A}(t)$ and $c^{\rm B}(t)$ to capture a sensor's transient response along with its ability to recover from a previous high-concentration exposure}
\label{fig:protocols}
\end{figure}

Using the above characterization, we propose a benchmark to capture a sensor's transient sensory capability as well as a its ability to recover from a previous exposure to high/low concentration. To illustrate the need for recovery, consider a sensor that binds strongly to ligands. After being exposed to a high concentration of ligands, it may be poisoned (i.e., stuck at the bound state) and it experiences a low binding frequency $f_{\rm bd}$. To increase $f_{\rm bd}$ and the sensory capability, a recovery in a low concentration environment may be needed. It would restore the unbound state and allow for higher $f_{\rm bd}$ at the next concentration up-shift. 

To design a benchmark that simultaneously captures the two sensory capabilities, we introduce the step-wise protocols $c^{\rm A}(t)$ and $c^{\rm B}(t)$ as illustrated by \cref{fig:protocols}. 
In the benchmark, both protocols start by initializing the sensor in a high concentration $c_h$ to mimic the effect caused by a previous exposure to high concentration of ligands. Then at time $t=-t_{\rm reset}$, both the protocols lower the ligand concentration to $c_l$ for the sensor's recovery. Ultimately, the next up-shift signal starts at time $t=0$, when the two protocols start to differ: $c^{\rm A}(t>0)=c_h$ and $c^{\rm B}(t>0)=c_l$. Under this protocol the sensor's trajectory KL divergence starts to accumulate with positive constant weighting factor $F^{\rm AB}$ and $F^{\rm BA}$ as shown in \cref{fig:protocols}.

\begin{figure}
\includegraphics{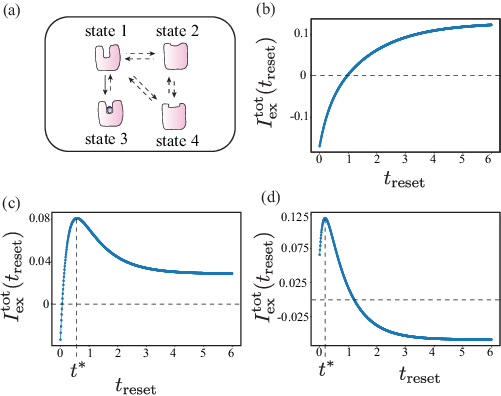}
\caption {(a) A 4-state sensor with one binding transition and 3 unbound states. (b) Sensor's performance improving as $t_{\rm reset}$ increases, as expected by intuition. (c) and (d) sensor performance exhibits non-monotonicity with respect to $t_{\rm reset}$ indicating a boost of performance at specific values of $t_{\rm reset}$.}
\label{fig:i_ex_example}
\end{figure}

{\it Performance function} $I^{\rm tot}_{\rm ex}(t_{\rm reset})$. With the proposed benchmark protocols, any ligand-receptor sensor's transient sensory capability $I^{\rm tot}_{\rm ex}(t_{\rm reset})$ and its dependence on the length of resetting time $t_{\rm reset}$ can be studied. This function characterizes the performance of the sensor by capturing both the transient sensory capability and its speed of recovery under resetting periods. A few example performance functions obtained from three different designs of four-state sensors are illustrated in \cref{fig:i_ex_example}a,b,c. 

Intuitively, one may expect a sensor's performance function $I^{\rm tot}_{\rm ex}(t_{\rm reset})$ to resemble the curve  shown in \cref{fig:i_ex_example}b. In this case, the transient sensory performance $I^{\rm tot}_{\rm ex}(0)$ under infinitely short recovery period is lower than that under an infinitely long recovery period ($I^{\rm tot}_{\rm ex}(\infty)\approx I^{\rm tot}_{\rm ex}(\tau_{\rm ss})$). In other words, the recovery helps the sensor resume the ability to facilitate binding transitions and thus helps the sensor achieve a better transient sensory performance. 

For illustrative purposes we denote the performances at the two ends of the performance function as
\begin{align}
    I^{\rm tot}_{{\rm ex},c_h}& \equiv I^{\rm tot}_{\rm ex}(0) \\
    I^{\rm tot}_{{\rm ex},c_l}& \equiv I^{\rm tot}_{\rm ex}(\tau_{\rm ss})
\end{align}
where the first one captures a sensor's transient sensory performance when the sensor is initiated at a high-concentration environment ($c_h$); and the second one captures the sensor's transient sensory capacity if the sensor is initialized at the steady state of a low concentration environment $c_l$, or equivalently, the sensor is fully recovered at a low concentration environment $c_l$. Notice that for some sensors, one may observe that its transient sensory performance initiated at high concentration may be better than that of the low concentration, $I^{\rm tot}_{{\rm ex},c_h}>I^{\rm tot}_{{\rm ex},c_l}$, as shown in \cref{fig:i_ex_example}(d). 

Notably, for some sensors the performance function is non-monotonic with respect to the recovery time, as shown in \cref{fig:i_ex_example}(c-d). For these sensors, the curves indicate that the sensors reach a boosted performance for specific $t_{\rm reset}$ values hence uncovering a design principle that can guide sensor performances to their best achievable values. This interesting behavior -- the sensory withdrawal effect -- of the sensor's transient sensory performance function is a result of the non-stationary dynamics of the sensor traversing different paths of its internal states and is discussed in detail below.

\section{Results}
\subsection{Markovian origin of sensory withdrawal effect}
One central observation of this work is a counter-intuitive withdrawal effect in sensory resetting: a previous exposure to high concentration of ligands followed by a brief period $t^*$ of low concentration can boost a sensor's sensitivity beyond any steady-state behavior. Specifically, for the same sensor under the same long observation time $\tau \gg\tau_{\rm ss}$, the ability for a sensor to distinguish an up-shift ligand concentration becomes the highest under an optimal recovery period $t^*<\tau_{\rm ss}$ \cref{fig:i_ex_example}b-d. Since the recovery does not impact $\dot D_{\rm ss}$, this effect can be alternatively characterized in terms of the total excess information: for an optimal resetting period $t^*$, both $I^{\rm tot}_{{\rm ex}}(t^*)>I^{\rm tot}_{{\rm ex},c_h}$ and $I^{\rm tot}_{{\rm ex}}(t^*)>I^{\rm tot}_{{\rm ex},c_l}$ are satisfied, as shown in \cref{fig:i_ex_example}c,d. This effect is a transient non-equilibrium effect due to the time-evolution of the system over the complex state space, and can not be explained by traditional steady-state analysis.

This withdrawal effect resembles another non-monotonic effect found in cooling/heating processes -- the Mpemba effect. The Mpemba effect describes that certain systems cool faster if they were previously heated to a higher initial temperature \cite{mpemba1969cool, lu2017nonequilibrium,  klich2019mpemba, chittari2023geometric}. It was shown that the Mpemba effect can be explained by decomposing the temporal evolution of the system into the combination of different eigenmodes. In the decomposition, it can be shown that the cooling time is dictated by the slowest eigenmode, where each mode relaxes under a exponential relaxation at a given rate (determined by the corresponding eigenvalue). Mpemba effect occurs if the decomposition factor of the slowest relaxation mode non-monotonically depends on the initial temperature. 
\begin{figure}
\includegraphics{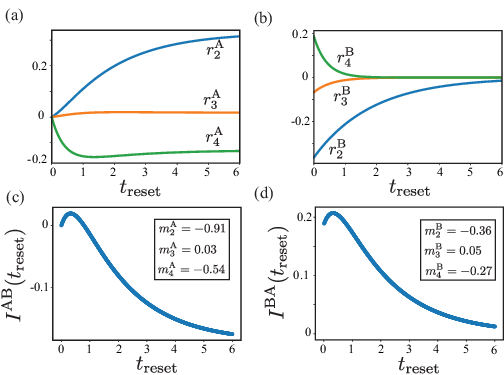}
\caption {(a) The coefficients $r^{\rm A}_i(t_{\rm reset})$, obtained for eigen decomposition of the sensor's initial state at $t=0$ into the eigenmodes of the sensor dynamics at a constant ligand concentration $c_h$. (b) The coefficients $r^{\rm B}_i(t_{\rm reset})$, obtained for eigen decomposition of the sensor's initial state at $t=0$ into the eigenmodes of the sensor dynamics at a constant ligand concentration $c_l$. (c) and (d) The respective $I^{\rm AB}(t_{\rm reset})$ and $I^{\rm AB}(t_{\rm reset})$ obtained using \cref{eq:eigenmodes} with the fixed contributions, $m_i^A$ and $m_i^B$ listed in the insets of (c) and (d). These results were obtained for the 4-state sensor shown in \cref{fig:i_ex_example}a with the following energy of each states: $E_1 = 0.8$, $E_2 = 0.4$, $E_3 = 0.8$, $E_4 = 0$, and barrier heights between the respective states: $B_{12}=2$, $B_{13}=1$, $B_{14}=1.1$, $B_{24}=0.5$ }.
\label{fig:coeff}
\end{figure}
Here we explain the non-monotonic withdrawal effect in terms of eigenmode decomposition. At time $t=0$, the sensor's state probability can be decomposed into a superposition of eigenmodes under the rate matrix corresponding to the constant concentration, $c_h$ or $c_l$. As proven in \cref{sec:level2}, each eigenmode ($i$) contributes a fixed total excess information $ m^{\rm A}_i$ and $ m^{\rm B}_i$:
\begin{align}
    I_{\rm ex}^{\rm tot}(t_{\rm reset}) & =I^{\rm AB}(t_{\rm reset})+I^{\rm BA}(t_{\rm reset})\\
    & =\sum_{i=2}^{N} [r^{\rm A}_i(t_{\rm reset}) \cdot m^{\rm A}_i +r^{\rm B}_i(t_{\rm reset})  \cdot m^{\rm B}_i]
    \label{eq:eigenmodes}
\end{align}
where the coefficients $r^{\rm A}_i(t_{\rm reset})$ as functions of the resetting time, are the decomposition factors of the sensor's initial state into the eigenmodes of the sensor dynamics at a constant ligand concentration $c_h$. Similarly, $r^{\rm B}_i(t_{\rm reset})$ are the decomposition of the sensor's initial state into the eigenmodes of the dynamics under concentration $c_l$. For a sensor with the withdrawal effect showing up in both $I^{\rm AB}(t_{\rm reset})$ and $I^{\rm BA}(t_{\rm reset})$ (see \cref{fig:coeff}c and \cref{fig:coeff}d), we find two different mechanisms behind the withdrawal effect. One mechanism resembles that found in the Markovian Mpemba effect, but the second mechanism is novel. In the first case, the withdrawal effect results from the non-monotonicity of the $4$-th eigenmode decomposition factor $r^{\rm A}_4(t_{\rm reset})$ as shown in \cref{fig:coeff}a. In the second case, all eigen decomposition factors are monotonic in time, as shown in \cref{fig:coeff}b, and the withdrawal effect is a result of the weighted summation of the three monotonic functions. This analysis reveals that the withdrawal effect is a result of more complex dynamical behaviors than the traditional explanation of the Mpemba effect, as it involves the contribution from multiple eigenmodes. 

\subsection{Structure origin of sensory withdrawal effect}
\begin{figure}
\includegraphics{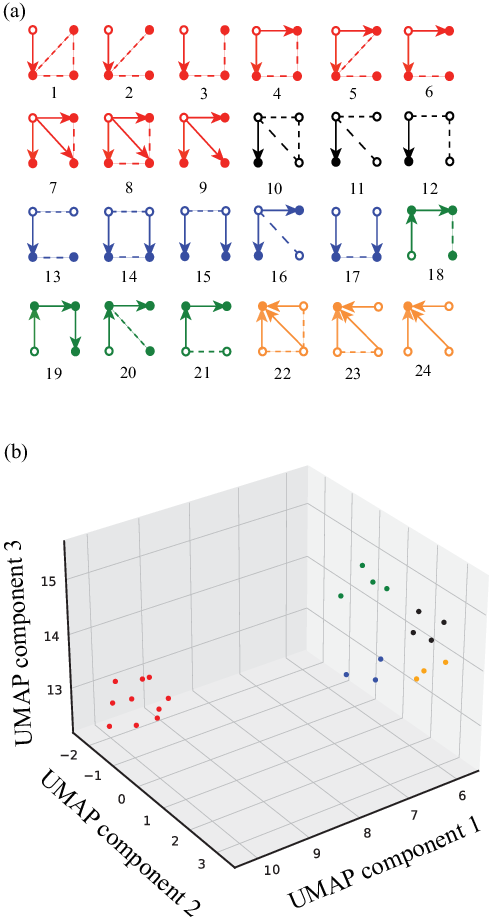}
\caption {4 state sensors classified into 5 classes based on the features of their graphs. (a) All the 4 state graphs considered in this study. A solid circle represents a bound state and a hollow circle represents an unbound state. The solid arrows represent a binding transition and the dotted lines represent non-binding transitions between states. Note that all the transitions are bi-directional. Each graph is colored based on the UMAP cluster as shown in (b).}
\label{fig:classification_graph}
\end{figure}
In the following, we utilize machine learning classification method to systematically investigate the connection between the sensor's state transition graph and the sensory withdrawal effect. Since the distinguishability and the excess information described previously is directly proportional to the average number of binding transitions, we expect sensors with multiple binding transitions to result in better performances.

To characterize the connection between the graph structure and the withdrawal effect, we estimate the frequencies of observing the withdrawal effect in 500 randomly generated energy landscapes combined with each possible 4-state-sensor graph shown in \cref{fig:classification_graph}a. In \cref{fig:classification_graph} and \cref{tab:table1}, we confirm a strong correlation between the existence of the withdrawal effect and the structural feature of the sensor's state graph. 

\begin{table}
\caption{\label{tab:table1}
For each graph in  \cref{fig:classification_graph}(a) the percentage of occurance of withdrawal effect}
\begin{ruledtabular}
\begin{tabular}{cc@{\hskip 1.5cm}|@{\hskip 0.1cm}cc} 
 Graph & $w\%$ & Graph & $w\%$ \\
\hline
\color{red}1 & \color{red}0 & \color{blue}13 & \color{blue}1.4\\ 
 \color{red}2 & \color{red}0 & \color{blue}14 & \color{blue}2.8 \\
 \color{red}3 & \color{red}0 & \color{blue}15 & \color{blue}14.6 \\
 \color{red}4 & \color{red}0 & \color{blue}16 & \color{blue}42\\
 \color{red}5 & \color{red}0 & \color{blue}17 & \color{blue}60.2\\ 
 \color{red}6 & \color{red}0 & \color{ForestGreen}18 & \color{ForestGreen}0\\
 \color{red}7 & \color{red}0 & \color{ForestGreen}19 & \color{ForestGreen}14\\
 \color{red}8 & \color{red}0 & \color{ForestGreen}20 & \color{ForestGreen}19.8\\
 \color{red}9 & \color{red}0 & \color{ForestGreen}21 & \color{ForestGreen}48.8\\
 10 & 51.4 & \color{orange}22 & \color{orange}76.8\\
 11 & 63.8 & \color{orange}23 & \color{orange}98.6\\
 12 & 96.8 & \color{orange}24 & \color{orange}100
\end{tabular}
\label{table:1}
\end{ruledtabular}
\end{table}

\begin{table*}[h!]
\caption{\label{tab:table2} Correlation between graph structure and withdrawal effect.}
\begin{ruledtabular}
\begin{tabular}{cccc}
 Class & Withdrawal effect & Graph features & Interpretation\\ \hline
 {1} & No & \makecell[l]{
      1 unbound state, 3 bound states.\\
     There could be 1, 2 or 3 \\ binding transitions.} & \makecell[l]{Since there is only one unbound state the recovery dynamics is \\too simple to allow for a withdrawal effect.} \\ 
  \hline
  2 & Yes & \makecell[l]{3 unbound states, 1 bound state.\\Only one unbound state\\ is allowed to bind.} & \makecell[l]{Multiple unbound states with only one unbound state \\allowed for the binding transition. These graphs allow \\for complex dynamics under the recovery period and thus \\allow for the withdrawal effect. The less transitions \\between the unbound states, the stronger the non-monotonicity of \\its resetting relaxation and thus higher the \\probability to see the withdrawal effect.}\\ 
  \hline
  3 & Yes & \makecell[l]{2 unbound states, 2 bound states.} & \makecell[l]{The stronger the connection (direct or indirect) between the two\\ unbound states, the weaker the withdrawal effect in agreement\\ with class 2.}\\
  
  \hline
  4 & Yes except 1 & \makecell[l]{The only class of graphs that \\allows for multiple levels of\\ binding ($k>1$).} & \makecell[l]{No clear trend is observed possibly due to the\\ multiplicity of the bound states ($k>1$).}\\
  
  \hline
  5 & Yes & \makecell[l]{3 unbound states, 1 bound state. \\ All unbound states can make \\binding transitions.} & \makecell[l]{
  In agreement with the observation from class 2 and 3, the less \\connection between the unbound states the more non-monotonicity\\ one could achieve in the recovery period thus allowing for more \\probability to observe the withdrawal effect. }\\
\end{tabular}
\end{ruledtabular}
\end{table*}

To capture the structure of the Markov graph we use the following features -- `the number of unbound states', `the number of unbound to unbound transitions', `the number of unbound states capable of binding transition', `the number of singly bound states not capable of binding', `the number of singly bound states capable of binding', `the number of singly-bound to singly-bound transitions' -- to classify the graphs with  Uniform Manifold Approximation and Projection (UMAP) embeddings in 3 dimensions \cite{mcinnes2018umap}. The clustering of the UMAP embeddings is shown in \cref{fig:classification_graph}(b). 

The classification of the sensor graph has strong correlation with the occurance of the sensory withdrawal effect. Graphs in classes 2-5 show withdrawal effect with one exception (in class 4) and the graphs in class 1 do not show withdrawal effect. The features of the graphs in each class and their interpretation are elaborated in \cref{tab:table2}. 

\subsection{Correlation between graph features and performance}
\begin{figure}
\includegraphics{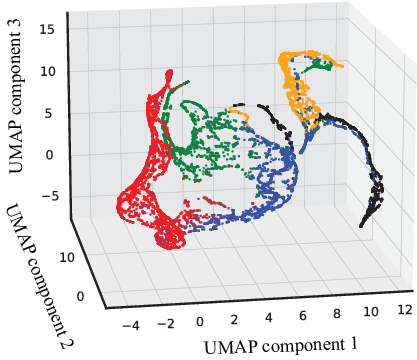}
\caption {The clustering of the UMAP embeddings obtained for the performance of sensors show the same 5 classes as obtained for the graph features confirming a strong correlation between graph features and performance of the sensor.}
\label{fig:i-umap}
\end{figure}

To systematically investigate the relationship between the sensor's performance and the structure of its state graph, we employ UMAP to visualize the high-dimensional performance data in a 3D latent space. The UMAP representation is generated from four-dimensional vectors capturing key features of the sensor's performance function, $I_{\rm ex}^{\rm tot}(t_{\rm reset})$. These features include the position and value of the maximum $I_{\rm ex}^{\rm tot}(t_{\rm reset})$ within the range (0,1), as well as the ratios of $I_{\rm ex}^{\rm tot}$ at the initial $(t_{\rm reset}=0)$ and final $(t_{\rm reset}^{\rm max})$ reset times to the maximum value of $I_{\rm ex}^{\rm tot}(t_{\rm reset})$. Remarkably, the UMAP embeddings of the sensor performances form distinct clusters that precisely match the previously identified graph classifications (\cref{fig:classification_graph}), as illustrated by the consistent color-coding in \cref{fig:i-umap}. This agreement between the performance-based clustering and the graph classification suggests an inherent link between the sensor's performance characteristics and the structural features of its underlying state graph.

\begin{figure*}
\includegraphics{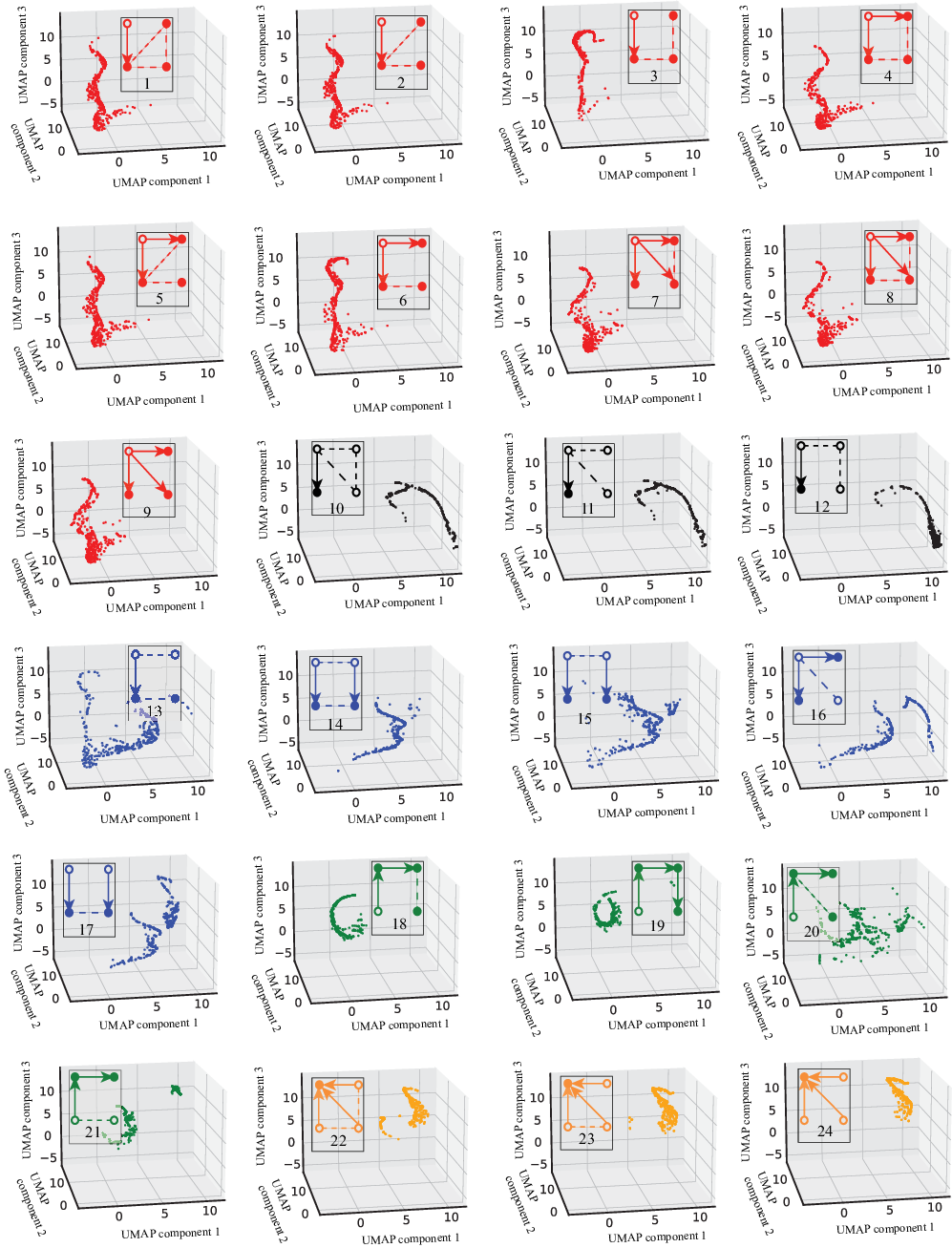}
\caption {UMAP embeddings calculated for the performance of all 24 kinds of sensors for 500 energy landscapes each. }
\label{fig:24_seperate}
\end{figure*}

To better visualize the correlation between the graph features and the performances of the sensor beyond the withdrawal effect, we plot the performance embeddings separately for each graph in \cref{fig:24_seperate}. Clearly, the performance embedding from graphs of the same structural class show similar shapes in the 3d latent space. It indicates that the structure of the Markov graph is tightly connected to the information performance of a sensor.

\section{Conclusions}
In this paper, we have provided a theoretical framework and a systematic benchmark to study the transient information sensory performance of ligand receptor sensor. We have identified a counter intuitive transient sensory withdrawal effect; in this effect, a sensor initiated in a high concentration environment followed by a brief recovery at low concentration can show significant boost in it's sensory performance in comparison to starting at any steady state. 

We confirm that the structure of the Markov graph of the sensor's states is tightly connected to the withdrawal effect and moreover the transient information performance over all. This is confirmed by the agreement of the classification of the sensors in both the structural embeddings and in the performance embeddings. This work provides a structural intuition for designing sensors with desired transient sensory performances. 

The numerical codes written for this study in Julia and Python languages will be made publicly available on GitHub. 

\section{ACKNOWLEDGMENTS} The authors appreciate financial support by funds from the National Science Foundation Grant No. DMR-2145256. We would like to thank the University of North Carolina at Chapel Hill and the Research Computing group for providing computational resources and support that have contributed to these research results.

\appendix
\section{\label{sec:level1}Ligand-receptor sensor's performance expressed by binding frequencies.}
For any transient process that is arbitrarily far from the steady state, the trajectory KL divergence can be expressed as an accumulated weighted sum of all observed transition events  \cite{pagare2024stochastic}:
\begin{equation}
    D_{{\rm KL}}^{\rm AB}(\tau) = \sum_{\langle x, x'\rangle} \int_0^ \tau    J^{\rm A}_{x'x}(t) \cdot F^{\rm AB}_{x'x}(t)~{\rm d} t
    \label{sieq:simu}
\end{equation}
where $J^{\rm A}_{x'x}(t)=R^{A}_{x'x}(t) \cdot p^{\rm A}(x;t)$ is the detailed probability current for sensor's transition from state $x$ to $x'$ at time $t$ under signal protocol ${\rm A}$; the weighting factor $F_{x' x}^{\rm AB}(t)$ characterizes the transition rate difference for each transition at time $t$ for the sensor's dynamics under the two signals. In this description, we have adopted the master equation description of the sensor's dynamics, where the transition probability rate from state $x$ to $x'$ is denoted by $R^{A}_{x'x}(t)$ and the probability of the sensor being in state $x$ at time $t$ under the protocol ${\rm A}$ is denoted by $p^{\rm A}(x;t)$. For ligand receptor sensors, the external signal is ligand concentration, and it only impacts the probability rates of the binding transitions. Thus one can show that the non-binding transitions do not contribute to the KL divergence (i.e., with a zero weighting factor as shown in the Eq.~10 of reference \cite{pagare2024stochastic}). Furthermore, we can show that for any binding transition, the weighting factor follows a simple formula as shown in \cref{eq:weighting_factor}). Ultimately, one can show that for ligand-receptor sensors, the KL divergence is simply a time integral of all binding frequencies weighted by the ``transient concentration difference'' between the two signal protocols, as shown in \cref{eq:KL_sensor}.

\section{\label{sec:level2}Eigen-mode decomposition of the total excess information}

By using the eigen representation of the sensor's state probability, one can express the initial probability $p^{\rm A}(x, 0;t_{\rm reset}) $ as the weighted sums of the eigenvectors of the rate matrix $R^{\rm A}$. 
\begin{eqnarray}
    p(x, t=0;t_{\rm reset}) = \pi^{\rm A}(x) + \sum_{i \ge 2} r_i^{\rm A}(t_{\rm reset}) v_i^{\rm A}(x, t)
\end{eqnarray}
Here the weighting factors are collectively expressed by a vector $r^{\rm A}(t_{\rm reset})$, and the leading eigenvector given by the steady-state distribution ${\pi}^{\rm A}$. Then as the sensor evolves according to the protocol $\rm A$, each of its eigenmode exhibits an exponential decay with the decay rate set by the corresponding eigenvalues:
\begin{eqnarray}
    p^{\rm A}(x, t) = \pi^{\rm A}(x) + \sum_{i \ge 2} r_i^{\rm A}(t_{\rm reset}) e^{\lambda_i^{\rm A} t}v_i^{\rm A}(x, t),
\end{eqnarray}
where $\pi^{\rm A}(x)$ is the steady-state distribution on state $x$, $\lambda_i$ is the $i$-th eigenvalue, and $v_i(x, t)$ is the component of the $i$-th eigenvector of matrix $R^{\rm A}$ on state $x$. Here, the eigenvalues $\{\lambda_i\}$ are ranged in the order $0 = \lambda_1 > \lambda_2\ > \lambda_3 \geq \cdots \ge \lambda_N$.

The KL divergence in \cref{sieq:simu} then becomes
\begin{widetext}
\begin{equation}
    D^{\rm AB}_{\text{KL}}(\tau;t_{\rm reset}) = \int_0^\tau\mathrm{d}t \sum_{\text{edge }x\to x'} R^{\rm A}_{x'x} \pi^A(x) F^{\rm AB}(t) + \sum_{i \ge 2} r_i^{\rm A} (t_{\rm reset})\int_0^\tau\mathrm{d}t \sum_{\text{edge }x\to x'} R_{x'x}^{\rm A} e^{\lambda_i^{\rm A} t} v_i^{\rm A}(x, t) F^{\rm AB}(t),
    \label{eq: Dkl decomposition on p}
\end{equation}
\end{widetext}
where the KL divergence is linearly dependent on the coefficients $\{r_i^{\rm A}\}$.
For a time-independent rate matrix $R^{\rm A}$, only the term $e^{\lambda_i^{\rm A} t}$ in \cref{eq: Dkl decomposition on p} includes time $t$. In this case, \cref{eq: Dkl decomposition on p} reduces to the following form:
\begin{widetext}
\begin{align}
    D^{\rm AB}_{\text{KL}}(\tau; t_{\rm reset}) &= \sum_{\text{edge }x\to x'} R^{\rm A}_{x'x} \pi^{\rm A}(x) F^{\rm AB} \tau  + \sum_{i \ge 2} \left[ \frac{r_i^{\rm A}(t_{\rm reset})}{\lambda_i^{\rm A}} (e^{\lambda_i^{\rm A} \tau}-1) \sum_{\text{edge }x\to x'} R_{x'x}^{\rm A} v_i^{\rm A}(x) F^{\rm AB} \right]\\
    &= \tau \cdot (f^{\rm {\rm A}}_{\rm ss}F^{\rm AB}) + \sum_{i \ge 2} \left[ \frac{r_i^{\rm A}(t_{\rm reset})}{\lambda_i^{\rm A}} (e^{\lambda_i^{\rm A} \tau}-1) \sum_{\text{edge }x\to x'} R_{x'x}^{\rm A} v_i^{\rm A}(x) F^{\rm AB} \right]
    \label{eq:DKL_final}
\end{align}
\end{widetext}

We used a symmetric version of the KL divergence $D_{\rm KL}(\tau) = D^{ \rm AB}_{\rm KL}(\tau)+D^{ \rm BA}_{\rm KL}(\tau)$ to define the sensor's transient sensory capability in terms of it's total excess information, $I^{\rm tot}_{\rm ex}(t_{\rm reset})$ and its dependence on the length of resetting time $t_{\rm reset}$:
\begin{equation}
    I_{\rm ex}^{\rm tot}(t_{\rm reset}) = \lim_{\tau\rightarrow \infty} D_{\rm KL}(\tau; t_{\rm reset})- D_{\rm ss}(\tau; t_{\rm reset})
    \label{eq:I_tot}
\end{equation}
Using \cref{eq:DKL_final}, \cref{eq:I_tot} for a $N$ state sensor becomes
\begin{widetext}
\begin{align}
    I_{\rm ex}^{\rm tot}(t_{\rm reset}) = \sum_{i \ge 2} \left[ \frac{-r_i^{\rm A}(t_{\rm reset})}{\lambda_i^{\rm A}} \sum_{\text{edge }x\to x'} R_{x'x}^{\rm A} v_i^{\rm A}(x) F^{\rm AB} + \frac{-r_i^{\rm B}(t_{\rm reset})}{\lambda_i^{\rm B}} \sum_{\text{edge }x\to x'} R_{x'x}^{\rm B} v_i^{\rm B}(x) F^{\rm BA} \right]
    \label{eq:i_tot_2}
\end{align}
\end{widetext}
which gives \cref{eq:eigenmodes} in the main text:
\begin{align}
    I_{\rm ex}^{\rm tot}(t_{\rm reset})
    & =\sum_{i\ge 2} [r^{\rm A}_i(t_{\rm reset}) \cdot m^{\rm A}_i +r^{\rm B}_i(t_{\rm reset})  \cdot m^{\rm B}_i]
\end{align}
where 
\begin{align}
    m^{\rm A}_i = -\frac{1}{\lambda_i^{\rm A}}R_{x'x}^{\rm A} v_i^{\rm A}(x) F^{\rm AB} \\
    m^{\rm B}_i = -\frac{1}{\lambda_i^{\rm B}}R_{x'x}^{\rm B} v_i^{\rm B}(x) F^{\rm BA}
\end{align}

\bibliography{apssamp}

\providecommand{\noopsort}[1]{}\providecommand{\singleletter}[1]{#1}%
\begin{thebibliography}{46}%
\makeatletter
\providecommand \@ifxundefined [1]{%
 \@ifx{#1\undefined}
}%
\providecommand \@ifnum [1]{%
 \ifnum #1\expandafter \@firstoftwo
 \else \expandafter \@secondoftwo
 \fi
}%
\providecommand \@ifx [1]{%
 \ifx #1\expandafter \@firstoftwo
 \else \expandafter \@secondoftwo
 \fi
}%
\providecommand \natexlab [1]{#1}%
\providecommand \enquote  [1]{``#1''}%
\providecommand \bibnamefont  [1]{#1}%
\providecommand \bibfnamefont [1]{#1}%
\providecommand \citenamefont [1]{#1}%
\providecommand \href@noop [0]{\@secondoftwo}%
\providecommand \href [0]{\begingroup \@sanitize@url \@href}%
\providecommand \@href[1]{\@@startlink{#1}\@@href}%
\providecommand \@@href[1]{\endgroup#1\@@endlink}%
\providecommand \@sanitize@url [0]{\catcode `\\12\catcode `\$12\catcode `\&12\catcode `\#12\catcode `\^12\catcode `\_12\catcode `\%12\relax}%
\providecommand \@@startlink[1]{}%
\providecommand \@@endlink[0]{}%
\providecommand \url  [0]{\begingroup\@sanitize@url \@url }%
\providecommand \@url [1]{\endgroup\@href {#1}{\urlprefix }}%
\providecommand \urlprefix  [0]{URL }%
\providecommand \Eprint [0]{\href }%
\providecommand \doibase [0]{https://doi.org/}%
\providecommand \selectlanguage [0]{\@gobble}%
\providecommand \bibinfo  [0]{\@secondoftwo}%
\providecommand \bibfield  [0]{\@secondoftwo}%
\providecommand \translation [1]{[#1]}%
\providecommand \BibitemOpen [0]{}%
\providecommand \bibitemStop [0]{}%
\providecommand \bibitemNoStop [0]{.\EOS\space}%
\providecommand \EOS [0]{\spacefactor3000\relax}%
\providecommand \BibitemShut  [1]{\csname bibitem#1\endcsname}%
\let\auto@bib@innerbib\@empty
\bibitem [{\citenamefont {Antebi}\ \emph {et~al.}(2017{\natexlab{a}})\citenamefont {Antebi}, \citenamefont {Linton}, \citenamefont {Klumpe}, \citenamefont {Bintu}, \citenamefont {Gong}, \citenamefont {Su}, \citenamefont {McCardell},\ and\ \citenamefont {Elowitz}}]{antebi2017combinatorial}%
  \BibitemOpen
  \bibfield  {author} {\bibinfo {author} {\bibfnamefont {Y.~E.}\ \bibnamefont {Antebi}}, \bibinfo {author} {\bibfnamefont {J.~M.}\ \bibnamefont {Linton}}, \bibinfo {author} {\bibfnamefont {H.}~\bibnamefont {Klumpe}}, \bibinfo {author} {\bibfnamefont {B.}~\bibnamefont {Bintu}}, \bibinfo {author} {\bibfnamefont {M.}~\bibnamefont {Gong}}, \bibinfo {author} {\bibfnamefont {C.}~\bibnamefont {Su}}, \bibinfo {author} {\bibfnamefont {R.}~\bibnamefont {McCardell}},\ and\ \bibinfo {author} {\bibfnamefont {M.~B.}\ \bibnamefont {Elowitz}},\ }\bibfield  {title} {\bibinfo {title} {Combinatorial signal perception in the bmp pathway},\ }\href@noop {} {\bibfield  {journal} {\bibinfo  {journal} {Cell}\ }\textbf {\bibinfo {volume} {170}},\ \bibinfo {pages} {1184} (\bibinfo {year} {2017}{\natexlab{a}})}\BibitemShut {NoStop}%
\bibitem [{\citenamefont {Antebi}\ \emph {et~al.}(2017{\natexlab{b}})\citenamefont {Antebi}, \citenamefont {Nandagopal},\ and\ \citenamefont {Elowitz}}]{antebi2017operational}%
  \BibitemOpen
  \bibfield  {author} {\bibinfo {author} {\bibfnamefont {Y.~E.}\ \bibnamefont {Antebi}}, \bibinfo {author} {\bibfnamefont {N.}~\bibnamefont {Nandagopal}},\ and\ \bibinfo {author} {\bibfnamefont {M.~B.}\ \bibnamefont {Elowitz}},\ }\bibfield  {title} {\bibinfo {title} {An operational view of intercellular signaling pathways},\ }\href@noop {} {\bibfield  {journal} {\bibinfo  {journal} {Current opinion in systems biology}\ }\textbf {\bibinfo {volume} {1}},\ \bibinfo {pages} {16} (\bibinfo {year} {2017}{\natexlab{b}})}\BibitemShut {NoStop}%
\bibitem [{\citenamefont {Lestas}\ \emph {et~al.}(2010)\citenamefont {Lestas}, \citenamefont {Vinnicombe},\ and\ \citenamefont {Paulsson}}]{lestas2010fundamental}%
  \BibitemOpen
  \bibfield  {author} {\bibinfo {author} {\bibfnamefont {I.}~\bibnamefont {Lestas}}, \bibinfo {author} {\bibfnamefont {G.}~\bibnamefont {Vinnicombe}},\ and\ \bibinfo {author} {\bibfnamefont {J.}~\bibnamefont {Paulsson}},\ }\bibfield  {title} {\bibinfo {title} {Fundamental limits on the suppression of molecular fluctuations},\ }\href@noop {} {\bibfield  {journal} {\bibinfo  {journal} {Nature}\ }\textbf {\bibinfo {volume} {467}},\ \bibinfo {pages} {174} (\bibinfo {year} {2010})}\BibitemShut {NoStop}%
\bibitem [{\citenamefont {Hinczewski}\ and\ \citenamefont {Thirumalai}(2014)}]{hinczewski2014cellular}%
  \BibitemOpen
  \bibfield  {author} {\bibinfo {author} {\bibfnamefont {M.}~\bibnamefont {Hinczewski}}\ and\ \bibinfo {author} {\bibfnamefont {D.}~\bibnamefont {Thirumalai}},\ }\bibfield  {title} {\bibinfo {title} {Cellular signaling networks function as generalized wiener-kolmogorov filters to suppress noise},\ }\href@noop {} {\bibfield  {journal} {\bibinfo  {journal} {Physical Review X}\ }\textbf {\bibinfo {volume} {4}},\ \bibinfo {pages} {041017} (\bibinfo {year} {2014})}\BibitemShut {NoStop}%
\bibitem [{\citenamefont {Berg}\ and\ \citenamefont {Purcell}(1977)}]{berg1977physics}%
  \BibitemOpen
  \bibfield  {author} {\bibinfo {author} {\bibfnamefont {H.~C.}\ \bibnamefont {Berg}}\ and\ \bibinfo {author} {\bibfnamefont {E.~M.}\ \bibnamefont {Purcell}},\ }\bibfield  {title} {\bibinfo {title} {Physics of chemoreception},\ }\href@noop {} {\bibfield  {journal} {\bibinfo  {journal} {Biophysical journal}\ }\textbf {\bibinfo {volume} {20}},\ \bibinfo {pages} {193} (\bibinfo {year} {1977})}\BibitemShut {NoStop}%
\bibitem [{\citenamefont {Bialek}\ and\ \citenamefont {Setayeshgar}(2005)}]{bialek2005physical}%
  \BibitemOpen
  \bibfield  {author} {\bibinfo {author} {\bibfnamefont {W.}~\bibnamefont {Bialek}}\ and\ \bibinfo {author} {\bibfnamefont {S.}~\bibnamefont {Setayeshgar}},\ }\bibfield  {title} {\bibinfo {title} {Physical limits to biochemical signaling},\ }\href@noop {} {\bibfield  {journal} {\bibinfo  {journal} {Proceedings of the National Academy of Sciences}\ }\textbf {\bibinfo {volume} {102}},\ \bibinfo {pages} {10040} (\bibinfo {year} {2005})}\BibitemShut {NoStop}%
\bibitem [{\citenamefont {Kaizu}\ \emph {et~al.}(2014)\citenamefont {Kaizu}, \citenamefont {De~Ronde}, \citenamefont {Paijmans}, \citenamefont {Takahashi}, \citenamefont {Tostevin},\ and\ \citenamefont {Ten~Wolde}}]{kaizu2014berg}%
  \BibitemOpen
  \bibfield  {author} {\bibinfo {author} {\bibfnamefont {K.}~\bibnamefont {Kaizu}}, \bibinfo {author} {\bibfnamefont {W.}~\bibnamefont {De~Ronde}}, \bibinfo {author} {\bibfnamefont {J.}~\bibnamefont {Paijmans}}, \bibinfo {author} {\bibfnamefont {K.}~\bibnamefont {Takahashi}}, \bibinfo {author} {\bibfnamefont {F.}~\bibnamefont {Tostevin}},\ and\ \bibinfo {author} {\bibfnamefont {P.~R.}\ \bibnamefont {Ten~Wolde}},\ }\bibfield  {title} {\bibinfo {title} {The berg-purcell limit revisited},\ }\href@noop {} {\bibfield  {journal} {\bibinfo  {journal} {Biophysical journal}\ }\textbf {\bibinfo {volume} {106}},\ \bibinfo {pages} {976} (\bibinfo {year} {2014})}\BibitemShut {NoStop}%
\bibitem [{\citenamefont {Singh}\ and\ \citenamefont {Nemenman}(2015)}]{singh2015accurate}%
  \BibitemOpen
  \bibfield  {author} {\bibinfo {author} {\bibfnamefont {V.}~\bibnamefont {Singh}}\ and\ \bibinfo {author} {\bibfnamefont {I.}~\bibnamefont {Nemenman}},\ }\bibfield  {title} {\bibinfo {title} {Accurate sensing of multiple ligands with a single receptor},\ }\href@noop {} {\bibfield  {journal} {\bibinfo  {journal} {arXiv preprint arXiv:1506.00288}\ } (\bibinfo {year} {2015})}\BibitemShut {NoStop}%
\bibitem [{\citenamefont {Nguyen}\ \emph {et~al.}(2015)\citenamefont {Nguyen}, \citenamefont {Dayan},\ and\ \citenamefont {Goodhill}}]{nguyen2015receptor}%
  \BibitemOpen
  \bibfield  {author} {\bibinfo {author} {\bibfnamefont {H.}~\bibnamefont {Nguyen}}, \bibinfo {author} {\bibfnamefont {P.}~\bibnamefont {Dayan}},\ and\ \bibinfo {author} {\bibfnamefont {G.}~\bibnamefont {Goodhill}},\ }\bibfield  {title} {\bibinfo {title} {How receptor diffusion influences gradient sensing},\ }\href@noop {} {\bibfield  {journal} {\bibinfo  {journal} {Journal of The Royal Society Interface}\ }\textbf {\bibinfo {volume} {12}},\ \bibinfo {pages} {20141097} (\bibinfo {year} {2015})}\BibitemShut {NoStop}%
\bibitem [{\citenamefont {Bialek}\ and\ \citenamefont {Setayeshgar}(2008)}]{bialek2008cooperativity}%
  \BibitemOpen
  \bibfield  {author} {\bibinfo {author} {\bibfnamefont {W.}~\bibnamefont {Bialek}}\ and\ \bibinfo {author} {\bibfnamefont {S.}~\bibnamefont {Setayeshgar}},\ }\bibfield  {title} {\bibinfo {title} {Cooperativity, sensitivity, and noise in biochemical signaling},\ }\href@noop {} {\bibfield  {journal} {\bibinfo  {journal} {Physical Review Letters}\ }\textbf {\bibinfo {volume} {100}},\ \bibinfo {pages} {258101} (\bibinfo {year} {2008})}\BibitemShut {NoStop}%
\bibitem [{\citenamefont {Bialek}(1997)}]{bialek1997statistical}%
  \BibitemOpen
  \bibfield  {author} {\bibinfo {author} {\bibfnamefont {W.}~\bibnamefont {Bialek}},\ }\bibfield  {title} {\bibinfo {title} {Statistical mechanics and sensory signal processing},\ }in\ \href@noop {} {\emph {\bibinfo {booktitle} {Physics of Biological Systems: From Molecules to Species}}}\ (\bibinfo {organization} {Springer},\ \bibinfo {year} {1997})\ pp.\ \bibinfo {pages} {252--280}\BibitemShut {NoStop}%
\bibitem [{\citenamefont {Marshall}(1995)}]{marshall1995specificity}%
  \BibitemOpen
  \bibfield  {author} {\bibinfo {author} {\bibfnamefont {C.}~\bibnamefont {Marshall}},\ }\bibfield  {title} {\bibinfo {title} {Specificity of receptor tyrosine kinase signaling: transient versus sustained extracellular signal-regulated kinase activation},\ }\href@noop {} {\bibfield  {journal} {\bibinfo  {journal} {Cell}\ }\textbf {\bibinfo {volume} {80}},\ \bibinfo {pages} {179} (\bibinfo {year} {1995})}\BibitemShut {NoStop}%
\bibitem [{\citenamefont {Hoffmann}\ \emph {et~al.}(2002)\citenamefont {Hoffmann}, \citenamefont {Levchenko}, \citenamefont {Scott},\ and\ \citenamefont {Baltimore}}]{hoffmann2002ikappab}%
  \BibitemOpen
  \bibfield  {author} {\bibinfo {author} {\bibfnamefont {A.}~\bibnamefont {Hoffmann}}, \bibinfo {author} {\bibfnamefont {A.}~\bibnamefont {Levchenko}}, \bibinfo {author} {\bibfnamefont {M.~L.}\ \bibnamefont {Scott}},\ and\ \bibinfo {author} {\bibfnamefont {D.}~\bibnamefont {Baltimore}},\ }\bibfield  {title} {\bibinfo {title} {The i$\kappa$b-nf-$\kappa$b signaling module: temporal control and selective gene activation},\ }\href@noop {} {\bibfield  {journal} {\bibinfo  {journal} {science}\ }\textbf {\bibinfo {volume} {298}},\ \bibinfo {pages} {1241} (\bibinfo {year} {2002})}\BibitemShut {NoStop}%
\bibitem [{\citenamefont {Purvis}\ and\ \citenamefont {Lahav}(2013)}]{purvis2013encoding}%
  \BibitemOpen
  \bibfield  {author} {\bibinfo {author} {\bibfnamefont {J.~E.}\ \bibnamefont {Purvis}}\ and\ \bibinfo {author} {\bibfnamefont {G.}~\bibnamefont {Lahav}},\ }\bibfield  {title} {\bibinfo {title} {Encoding and decoding cellular information through signaling dynamics},\ }\href@noop {} {\bibfield  {journal} {\bibinfo  {journal} {Cell}\ }\textbf {\bibinfo {volume} {152}},\ \bibinfo {pages} {945} (\bibinfo {year} {2013})}\BibitemShut {NoStop}%
\bibitem [{\citenamefont {Mora}\ and\ \citenamefont {Nemenman}(2019)}]{mora2019physical}%
  \BibitemOpen
  \bibfield  {author} {\bibinfo {author} {\bibfnamefont {T.}~\bibnamefont {Mora}}\ and\ \bibinfo {author} {\bibfnamefont {I.}~\bibnamefont {Nemenman}},\ }\bibfield  {title} {\bibinfo {title} {Physical limit to concentration sensing in a changing environment},\ }\href@noop {} {\bibfield  {journal} {\bibinfo  {journal} {Physical review letters}\ }\textbf {\bibinfo {volume} {123}},\ \bibinfo {pages} {198101} (\bibinfo {year} {2019})}\BibitemShut {NoStop}%
\bibitem [{\citenamefont {Hopfield}(1974)}]{hopfield1974kinetic}%
  \BibitemOpen
  \bibfield  {author} {\bibinfo {author} {\bibfnamefont {J.~J.}\ \bibnamefont {Hopfield}},\ }\bibfield  {title} {\bibinfo {title} {Kinetic proofreading: a new mechanism for reducing errors in biosynthetic processes requiring high specificity},\ }\href@noop {} {\bibfield  {journal} {\bibinfo  {journal} {Proceedings of the National Academy of Sciences}\ }\textbf {\bibinfo {volume} {71}},\ \bibinfo {pages} {4135} (\bibinfo {year} {1974})}\BibitemShut {NoStop}%
\bibitem [{\citenamefont {Behar}\ and\ \citenamefont {Hoffmann}(2010)}]{behar2010understanding}%
  \BibitemOpen
  \bibfield  {author} {\bibinfo {author} {\bibfnamefont {M.}~\bibnamefont {Behar}}\ and\ \bibinfo {author} {\bibfnamefont {A.}~\bibnamefont {Hoffmann}},\ }\bibfield  {title} {\bibinfo {title} {Understanding the temporal codes of intra-cellular signals},\ }\href@noop {} {\bibfield  {journal} {\bibinfo  {journal} {Current opinion in genetics \& development}\ }\textbf {\bibinfo {volume} {20}},\ \bibinfo {pages} {684} (\bibinfo {year} {2010})}\BibitemShut {NoStop}%
\bibitem [{\citenamefont {Qian}(2006)}]{qian2006reducing}%
  \BibitemOpen
  \bibfield  {author} {\bibinfo {author} {\bibfnamefont {H.}~\bibnamefont {Qian}},\ }\bibfield  {title} {\bibinfo {title} {Reducing intrinsic biochemical noise in cells and its thermodynamic limit},\ }\href@noop {} {\bibfield  {journal} {\bibinfo  {journal} {Journal of molecular biology}\ }\textbf {\bibinfo {volume} {362}},\ \bibinfo {pages} {387} (\bibinfo {year} {2006})}\BibitemShut {NoStop}%
\bibitem [{\citenamefont {Murugan}\ \emph {et~al.}(2012)\citenamefont {Murugan}, \citenamefont {Huse},\ and\ \citenamefont {Leibler}}]{murugan2012speed}%
  \BibitemOpen
  \bibfield  {author} {\bibinfo {author} {\bibfnamefont {A.}~\bibnamefont {Murugan}}, \bibinfo {author} {\bibfnamefont {D.~A.}\ \bibnamefont {Huse}},\ and\ \bibinfo {author} {\bibfnamefont {S.}~\bibnamefont {Leibler}},\ }\bibfield  {title} {\bibinfo {title} {Speed, dissipation, and error in kinetic proofreading},\ }\href@noop {} {\bibfield  {journal} {\bibinfo  {journal} {Proceedings of the National Academy of Sciences}\ }\textbf {\bibinfo {volume} {109}},\ \bibinfo {pages} {12034} (\bibinfo {year} {2012})}\BibitemShut {NoStop}%
\bibitem [{\citenamefont {Bauer}\ and\ \citenamefont {Bialek}(2023)}]{bauer2023information}%
  \BibitemOpen
  \bibfield  {author} {\bibinfo {author} {\bibfnamefont {M.}~\bibnamefont {Bauer}}\ and\ \bibinfo {author} {\bibfnamefont {W.}~\bibnamefont {Bialek}},\ }\bibfield  {title} {\bibinfo {title} {Information bottleneck in molecular sensing},\ }\href@noop {} {\bibfield  {journal} {\bibinfo  {journal} {PRX Life}\ }\textbf {\bibinfo {volume} {1}},\ \bibinfo {pages} {023005} (\bibinfo {year} {2023})}\BibitemShut {NoStop}%
\bibitem [{\citenamefont {Bialek}(2012)}]{bialek2012biophysics}%
  \BibitemOpen
  \bibfield  {author} {\bibinfo {author} {\bibfnamefont {W.}~\bibnamefont {Bialek}},\ }\href@noop {} {\emph {\bibinfo {title} {Biophysics: searching for principles}}}\ (\bibinfo  {publisher} {Princeton University Press},\ \bibinfo {year} {2012})\BibitemShut {NoStop}%
\bibitem [{\citenamefont {Potter}\ \emph {et~al.}(2017)\citenamefont {Potter}, \citenamefont {Byrd}, \citenamefont {Mugler},\ and\ \citenamefont {Sun}}]{potter2017dynamic}%
  \BibitemOpen
  \bibfield  {author} {\bibinfo {author} {\bibfnamefont {G.~D.}\ \bibnamefont {Potter}}, \bibinfo {author} {\bibfnamefont {T.~A.}\ \bibnamefont {Byrd}}, \bibinfo {author} {\bibfnamefont {A.}~\bibnamefont {Mugler}},\ and\ \bibinfo {author} {\bibfnamefont {B.}~\bibnamefont {Sun}},\ }\bibfield  {title} {\bibinfo {title} {Dynamic sampling and information encoding in biochemical networks},\ }\href@noop {} {\bibfield  {journal} {\bibinfo  {journal} {Biophysical journal}\ }\textbf {\bibinfo {volume} {112}},\ \bibinfo {pages} {795} (\bibinfo {year} {2017})}\BibitemShut {NoStop}%
\bibitem [{\citenamefont {Tang}\ \emph {et~al.}(2021)\citenamefont {Tang}, \citenamefont {Adelaja}, \citenamefont {Ye}, \citenamefont {Deeds}, \citenamefont {Wollman},\ and\ \citenamefont {Hoffmann}}]{tang2021quantifying}%
  \BibitemOpen
  \bibfield  {author} {\bibinfo {author} {\bibfnamefont {Y.}~\bibnamefont {Tang}}, \bibinfo {author} {\bibfnamefont {A.}~\bibnamefont {Adelaja}}, \bibinfo {author} {\bibfnamefont {F.~X.-F.}\ \bibnamefont {Ye}}, \bibinfo {author} {\bibfnamefont {E.}~\bibnamefont {Deeds}}, \bibinfo {author} {\bibfnamefont {R.}~\bibnamefont {Wollman}},\ and\ \bibinfo {author} {\bibfnamefont {A.}~\bibnamefont {Hoffmann}},\ }\bibfield  {title} {\bibinfo {title} {Quantifying information accumulation encoded in the dynamics of biochemical signaling},\ }\href@noop {} {\bibfield  {journal} {\bibinfo  {journal} {Nature communications}\ }\textbf {\bibinfo {volume} {12}},\ \bibinfo {pages} {1272} (\bibinfo {year} {2021})}\BibitemShut {NoStop}%
\bibitem [{\citenamefont {Cepeda-Humerez}\ \emph {et~al.}(2019)\citenamefont {Cepeda-Humerez}, \citenamefont {Ruess},\ and\ \citenamefont {Tka{\v{c}}ik}}]{cepeda2019estimating}%
  \BibitemOpen
  \bibfield  {author} {\bibinfo {author} {\bibfnamefont {S.~A.}\ \bibnamefont {Cepeda-Humerez}}, \bibinfo {author} {\bibfnamefont {J.}~\bibnamefont {Ruess}},\ and\ \bibinfo {author} {\bibfnamefont {G.}~\bibnamefont {Tka{\v{c}}ik}},\ }\bibfield  {title} {\bibinfo {title} {Estimating information in time-varying signals},\ }\href@noop {} {\bibfield  {journal} {\bibinfo  {journal} {PLoS computational biology}\ }\textbf {\bibinfo {volume} {15}},\ \bibinfo {pages} {e1007290} (\bibinfo {year} {2019})}\BibitemShut {NoStop}%
\bibitem [{\citenamefont {Hahn}\ \emph {et~al.}(2023)\citenamefont {Hahn}, \citenamefont {Walczak},\ and\ \citenamefont {Mora}}]{hahn2023dynamical}%
  \BibitemOpen
  \bibfield  {author} {\bibinfo {author} {\bibfnamefont {L.}~\bibnamefont {Hahn}}, \bibinfo {author} {\bibfnamefont {A.~M.}\ \bibnamefont {Walczak}},\ and\ \bibinfo {author} {\bibfnamefont {T.}~\bibnamefont {Mora}},\ }\bibfield  {title} {\bibinfo {title} {Dynamical information synergy in biochemical signaling networks},\ }\href@noop {} {\bibfield  {journal} {\bibinfo  {journal} {Physical Review Letters}\ }\textbf {\bibinfo {volume} {131}},\ \bibinfo {pages} {128401} (\bibinfo {year} {2023})}\BibitemShut {NoStop}%
\bibitem [{\citenamefont {Pagare}\ \emph {et~al.}(2023)\citenamefont {Pagare}, \citenamefont {Min},\ and\ \citenamefont {Lu}}]{pagare2023theoretical}%
  \BibitemOpen
  \bibfield  {author} {\bibinfo {author} {\bibfnamefont {A.}~\bibnamefont {Pagare}}, \bibinfo {author} {\bibfnamefont {S.~H.}\ \bibnamefont {Min}},\ and\ \bibinfo {author} {\bibfnamefont {Z.}~\bibnamefont {Lu}},\ }\bibfield  {title} {\bibinfo {title} {Theoretical upper bound of multiplexing in biological sensory receptors},\ }\href@noop {} {\bibfield  {journal} {\bibinfo  {journal} {Physical Review Research}\ }\textbf {\bibinfo {volume} {5}},\ \bibinfo {pages} {023032} (\bibinfo {year} {2023})}\BibitemShut {NoStop}%
\bibitem [{\citenamefont {Smith}\ \emph {et~al.}(2018)\citenamefont {Smith}, \citenamefont {Lefkowitz},\ and\ \citenamefont {Rajagopal}}]{smith2018biased}%
  \BibitemOpen
  \bibfield  {author} {\bibinfo {author} {\bibfnamefont {J.~S.}\ \bibnamefont {Smith}}, \bibinfo {author} {\bibfnamefont {R.~J.}\ \bibnamefont {Lefkowitz}},\ and\ \bibinfo {author} {\bibfnamefont {S.}~\bibnamefont {Rajagopal}},\ }\bibfield  {title} {\bibinfo {title} {Biased signalling: from simple switches to allosteric microprocessors},\ }\href@noop {} {\bibfield  {journal} {\bibinfo  {journal} {Nature reviews Drug discovery}\ }\textbf {\bibinfo {volume} {17}},\ \bibinfo {pages} {243} (\bibinfo {year} {2018})}\BibitemShut {NoStop}%
\bibitem [{\citenamefont {Gregorio}\ \emph {et~al.}(2017)\citenamefont {Gregorio}, \citenamefont {Masureel}, \citenamefont {Hilger}, \citenamefont {Terry}, \citenamefont {Juette}, \citenamefont {Zhao}, \citenamefont {Zhou}, \citenamefont {Perez-Aguilar}, \citenamefont {Hauge}, \citenamefont {Mathiasen} \emph {et~al.}}]{gregorio2017single}%
  \BibitemOpen
  \bibfield  {author} {\bibinfo {author} {\bibfnamefont {G.~G.}\ \bibnamefont {Gregorio}}, \bibinfo {author} {\bibfnamefont {M.}~\bibnamefont {Masureel}}, \bibinfo {author} {\bibfnamefont {D.}~\bibnamefont {Hilger}}, \bibinfo {author} {\bibfnamefont {D.~S.}\ \bibnamefont {Terry}}, \bibinfo {author} {\bibfnamefont {M.}~\bibnamefont {Juette}}, \bibinfo {author} {\bibfnamefont {H.}~\bibnamefont {Zhao}}, \bibinfo {author} {\bibfnamefont {Z.}~\bibnamefont {Zhou}}, \bibinfo {author} {\bibfnamefont {J.~M.}\ \bibnamefont {Perez-Aguilar}}, \bibinfo {author} {\bibfnamefont {M.}~\bibnamefont {Hauge}}, \bibinfo {author} {\bibfnamefont {S.}~\bibnamefont {Mathiasen}}, \emph {et~al.},\ }\bibfield  {title} {\bibinfo {title} {Single-molecule analysis of ligand efficacy in $\beta$2ar--g-protein activation},\ }\href@noop {} {\bibfield  {journal} {\bibinfo  {journal} {Nature}\ }\textbf {\bibinfo {volume} {547}},\ \bibinfo {pages} {68} (\bibinfo {year} {2017})}\BibitemShut {NoStop}%
\bibitem [{\citenamefont {Che}\ \emph {et~al.}(2020)\citenamefont {Che}, \citenamefont {English}, \citenamefont {Krumm}, \citenamefont {Kim}, \citenamefont {Pardon}, \citenamefont {Olsen}, \citenamefont {Wang}, \citenamefont {Zhang}, \citenamefont {Diberto}, \citenamefont {Sciaky} \emph {et~al.}}]{che2020nanobody}%
  \BibitemOpen
  \bibfield  {author} {\bibinfo {author} {\bibfnamefont {T.}~\bibnamefont {Che}}, \bibinfo {author} {\bibfnamefont {J.}~\bibnamefont {English}}, \bibinfo {author} {\bibfnamefont {B.}~\bibnamefont {Krumm}}, \bibinfo {author} {\bibfnamefont {K.}~\bibnamefont {Kim}}, \bibinfo {author} {\bibfnamefont {E.}~\bibnamefont {Pardon}}, \bibinfo {author} {\bibfnamefont {R.}~\bibnamefont {Olsen}}, \bibinfo {author} {\bibfnamefont {S.}~\bibnamefont {Wang}}, \bibinfo {author} {\bibfnamefont {S.}~\bibnamefont {Zhang}}, \bibinfo {author} {\bibfnamefont {J.}~\bibnamefont {Diberto}}, \bibinfo {author} {\bibfnamefont {N.}~\bibnamefont {Sciaky}}, \emph {et~al.},\ }\bibfield  {title} {\bibinfo {title} {Nanobody-enabled monitoring of kappa opioid receptor states. nat commun 11: 1145},\ }\href@noop {} {\bibfield  {journal} {\bibinfo  {journal} {doi. org/10.1038/s41467-020-14889-7}\ } (\bibinfo {year} {2020})}\BibitemShut {NoStop}%
\bibitem [{\citenamefont {Zhang}\ \emph {et~al.}(2023)\citenamefont {Zhang}, \citenamefont {R{\'o}zsa}, \citenamefont {Liang}, \citenamefont {Bushey}, \citenamefont {Wei}, \citenamefont {Zheng}, \citenamefont {Reep}, \citenamefont {Broussard}, \citenamefont {Tsang}, \citenamefont {Tsegaye} \emph {et~al.}}]{zhang2023fast}%
  \BibitemOpen
  \bibfield  {author} {\bibinfo {author} {\bibfnamefont {Y.}~\bibnamefont {Zhang}}, \bibinfo {author} {\bibfnamefont {M.}~\bibnamefont {R{\'o}zsa}}, \bibinfo {author} {\bibfnamefont {Y.}~\bibnamefont {Liang}}, \bibinfo {author} {\bibfnamefont {D.}~\bibnamefont {Bushey}}, \bibinfo {author} {\bibfnamefont {Z.}~\bibnamefont {Wei}}, \bibinfo {author} {\bibfnamefont {J.}~\bibnamefont {Zheng}}, \bibinfo {author} {\bibfnamefont {D.}~\bibnamefont {Reep}}, \bibinfo {author} {\bibfnamefont {G.~J.}\ \bibnamefont {Broussard}}, \bibinfo {author} {\bibfnamefont {A.}~\bibnamefont {Tsang}}, \bibinfo {author} {\bibfnamefont {G.}~\bibnamefont {Tsegaye}}, \emph {et~al.},\ }\bibfield  {title} {\bibinfo {title} {Fast and sensitive gcamp calcium indicators for imaging neural populations},\ }\href@noop {} {\bibfield  {journal} {\bibinfo  {journal} {Nature}\ }\textbf {\bibinfo {volume} {615}},\ \bibinfo {pages} {884} (\bibinfo {year} {2023})}\BibitemShut {NoStop}%
\bibitem [{\citenamefont {Thomas}\ and\ \citenamefont {Tamp{\'e}}(2023)}]{thomas2023structure}%
  \BibitemOpen
  \bibfield  {author} {\bibinfo {author} {\bibfnamefont {C.}~\bibnamefont {Thomas}}\ and\ \bibinfo {author} {\bibfnamefont {R.}~\bibnamefont {Tamp{\'e}}},\ }\bibfield  {title} {\bibinfo {title} {Structure and mechanism of immunoreceptors: New horizons in t cell and b cell receptor biology and beyond},\ }\href@noop {} {\bibfield  {journal} {\bibinfo  {journal} {Current Opinion in Structural Biology}\ }\textbf {\bibinfo {volume} {80}},\ \bibinfo {pages} {102570} (\bibinfo {year} {2023})}\BibitemShut {NoStop}%
\bibitem [{\citenamefont {Han}\ \emph {et~al.}(2023)\citenamefont {Han}, \citenamefont {Chen}, \citenamefont {Zhu},\ and\ \citenamefont {Huang}}]{han2023antigen}%
  \BibitemOpen
  \bibfield  {author} {\bibinfo {author} {\bibfnamefont {F.}~\bibnamefont {Han}}, \bibinfo {author} {\bibfnamefont {Y.}~\bibnamefont {Chen}}, \bibinfo {author} {\bibfnamefont {Y.}~\bibnamefont {Zhu}},\ and\ \bibinfo {author} {\bibfnamefont {Z.}~\bibnamefont {Huang}},\ }\bibfield  {title} {\bibinfo {title} {Antigen receptor structure and signaling},\ }\href@noop {} {\bibfield  {journal} {\bibinfo  {journal} {Advances in Immunology}\ }\textbf {\bibinfo {volume} {157}},\ \bibinfo {pages} {1} (\bibinfo {year} {2023})}\BibitemShut {NoStop}%
\bibitem [{\citenamefont {Shen}\ \emph {et~al.}(2019)\citenamefont {Shen}, \citenamefont {Liu}, \citenamefont {Li}, \citenamefont {Wan}, \citenamefont {Mao}, \citenamefont {Chen},\ and\ \citenamefont {Liu}}]{shen2019conformational}%
  \BibitemOpen
  \bibfield  {author} {\bibinfo {author} {\bibfnamefont {Z.}~\bibnamefont {Shen}}, \bibinfo {author} {\bibfnamefont {S.}~\bibnamefont {Liu}}, \bibinfo {author} {\bibfnamefont {X.}~\bibnamefont {Li}}, \bibinfo {author} {\bibfnamefont {Z.}~\bibnamefont {Wan}}, \bibinfo {author} {\bibfnamefont {Y.}~\bibnamefont {Mao}}, \bibinfo {author} {\bibfnamefont {C.}~\bibnamefont {Chen}},\ and\ \bibinfo {author} {\bibfnamefont {W.}~\bibnamefont {Liu}},\ }\bibfield  {title} {\bibinfo {title} {Conformational change within the extracellular domain of b cell receptor in b cell activation upon antigen binding},\ }\href@noop {} {\bibfield  {journal} {\bibinfo  {journal} {Elife}\ }\textbf {\bibinfo {volume} {8}},\ \bibinfo {pages} {e42271} (\bibinfo {year} {2019})}\BibitemShut {NoStop}%
\bibitem [{\citenamefont {De~Smet}\ \emph {et~al.}(2014)\citenamefont {De~Smet}, \citenamefont {Christopoulos},\ and\ \citenamefont {Carmeliet}}]{de2014allosteric}%
  \BibitemOpen
  \bibfield  {author} {\bibinfo {author} {\bibfnamefont {F.}~\bibnamefont {De~Smet}}, \bibinfo {author} {\bibfnamefont {A.}~\bibnamefont {Christopoulos}},\ and\ \bibinfo {author} {\bibfnamefont {P.}~\bibnamefont {Carmeliet}},\ }\bibfield  {title} {\bibinfo {title} {Allosteric targeting of receptor tyrosine kinases},\ }\href@noop {} {\bibfield  {journal} {\bibinfo  {journal} {Nature biotechnology}\ }\textbf {\bibinfo {volume} {32}},\ \bibinfo {pages} {1113} (\bibinfo {year} {2014})}\BibitemShut {NoStop}%
\bibitem [{\citenamefont {Chen}\ \emph {et~al.}(2017)\citenamefont {Chen}, \citenamefont {Marsiglia}, \citenamefont {Cho}, \citenamefont {Huang}, \citenamefont {Deng}, \citenamefont {Blais}, \citenamefont {Gai}, \citenamefont {Bhattacharya}, \citenamefont {Neubert}, \citenamefont {Traaseth} \emph {et~al.}}]{chen2017elucidation}%
  \BibitemOpen
  \bibfield  {author} {\bibinfo {author} {\bibfnamefont {H.}~\bibnamefont {Chen}}, \bibinfo {author} {\bibfnamefont {W.~M.}\ \bibnamefont {Marsiglia}}, \bibinfo {author} {\bibfnamefont {M.-K.}\ \bibnamefont {Cho}}, \bibinfo {author} {\bibfnamefont {Z.}~\bibnamefont {Huang}}, \bibinfo {author} {\bibfnamefont {J.}~\bibnamefont {Deng}}, \bibinfo {author} {\bibfnamefont {S.~P.}\ \bibnamefont {Blais}}, \bibinfo {author} {\bibfnamefont {W.}~\bibnamefont {Gai}}, \bibinfo {author} {\bibfnamefont {S.}~\bibnamefont {Bhattacharya}}, \bibinfo {author} {\bibfnamefont {T.~A.}\ \bibnamefont {Neubert}}, \bibinfo {author} {\bibfnamefont {N.~J.}\ \bibnamefont {Traaseth}}, \emph {et~al.},\ }\bibfield  {title} {\bibinfo {title} {Elucidation of a four-site allosteric network in fibroblast growth factor receptor tyrosine kinases},\ }\href@noop {} {\bibfield  {journal} {\bibinfo  {journal} {elife}\ }\textbf {\bibinfo {volume} {6}},\ \bibinfo {pages} {e21137} (\bibinfo {year} {2017})}\BibitemShut {NoStop}%
\bibitem [{\citenamefont {Hubbard}\ and\ \citenamefont {Miller}(2007)}]{hubbard2007receptor}%
  \BibitemOpen
  \bibfield  {author} {\bibinfo {author} {\bibfnamefont {S.~R.}\ \bibnamefont {Hubbard}}\ and\ \bibinfo {author} {\bibfnamefont {W.~T.}\ \bibnamefont {Miller}},\ }\bibfield  {title} {\bibinfo {title} {Receptor tyrosine kinases: mechanisms of activation and signaling},\ }\href@noop {} {\bibfield  {journal} {\bibinfo  {journal} {Current opinion in cell biology}\ }\textbf {\bibinfo {volume} {19}},\ \bibinfo {pages} {117} (\bibinfo {year} {2007})}\BibitemShut {NoStop}%
\bibitem [{\citenamefont {Volinsky}\ and\ \citenamefont {Kholodenko}(2013)}]{volinsky2013complexity}%
  \BibitemOpen
  \bibfield  {author} {\bibinfo {author} {\bibfnamefont {N.}~\bibnamefont {Volinsky}}\ and\ \bibinfo {author} {\bibfnamefont {B.~N.}\ \bibnamefont {Kholodenko}},\ }\bibfield  {title} {\bibinfo {title} {Complexity of receptor tyrosine kinase signal processing},\ }\href@noop {} {\bibfield  {journal} {\bibinfo  {journal} {Cold Spring Harbor perspectives in biology}\ }\textbf {\bibinfo {volume} {5}},\ \bibinfo {pages} {a009043} (\bibinfo {year} {2013})}\BibitemShut {NoStop}%
\bibitem [{\citenamefont {Sieghart}(2015)}]{sieghart2015allosteric}%
  \BibitemOpen
  \bibfield  {author} {\bibinfo {author} {\bibfnamefont {W.}~\bibnamefont {Sieghart}},\ }\bibfield  {title} {\bibinfo {title} {Allosteric modulation of gabaa receptors via multiple drug-binding sites},\ }\href@noop {} {\bibfield  {journal} {\bibinfo  {journal} {Advances in pharmacology}\ }\textbf {\bibinfo {volume} {72}},\ \bibinfo {pages} {53} (\bibinfo {year} {2015})}\BibitemShut {NoStop}%
\bibitem [{\citenamefont {May}\ \emph {et~al.}(2007)\citenamefont {May}, \citenamefont {Leach}, \citenamefont {Sexton},\ and\ \citenamefont {Christopoulos}}]{may2007allosteric}%
  \BibitemOpen
  \bibfield  {author} {\bibinfo {author} {\bibfnamefont {L.~T.}\ \bibnamefont {May}}, \bibinfo {author} {\bibfnamefont {K.}~\bibnamefont {Leach}}, \bibinfo {author} {\bibfnamefont {P.~M.}\ \bibnamefont {Sexton}},\ and\ \bibinfo {author} {\bibfnamefont {A.}~\bibnamefont {Christopoulos}},\ }\bibfield  {title} {\bibinfo {title} {Allosteric modulation of g protein--coupled receptors},\ }\href@noop {} {\bibfield  {journal} {\bibinfo  {journal} {Annu. Rev. Pharmacol. Toxicol.}\ }\textbf {\bibinfo {volume} {47}},\ \bibinfo {pages} {1} (\bibinfo {year} {2007})}\BibitemShut {NoStop}%
\bibitem [{\citenamefont {Deupi}\ and\ \citenamefont {Kobilka}(2010)}]{deupi2010energy}%
  \BibitemOpen
  \bibfield  {author} {\bibinfo {author} {\bibfnamefont {X.}~\bibnamefont {Deupi}}\ and\ \bibinfo {author} {\bibfnamefont {B.~K.}\ \bibnamefont {Kobilka}},\ }\bibfield  {title} {\bibinfo {title} {Energy landscapes as a tool to integrate gpcr structure, dynamics, and function},\ }\href@noop {} {\bibfield  {journal} {\bibinfo  {journal} {Physiology}\ }\textbf {\bibinfo {volume} {25}},\ \bibinfo {pages} {293} (\bibinfo {year} {2010})}\BibitemShut {NoStop}%
\bibitem [{\citenamefont {Pagare}\ \emph {et~al.}(2024)\citenamefont {Pagare}, \citenamefont {Zhang}, \citenamefont {Zheng},\ and\ \citenamefont {Lu}}]{pagare2024stochastic}%
  \BibitemOpen
  \bibfield  {author} {\bibinfo {author} {\bibfnamefont {A.}~\bibnamefont {Pagare}}, \bibinfo {author} {\bibfnamefont {Z.}~\bibnamefont {Zhang}}, \bibinfo {author} {\bibfnamefont {J.}~\bibnamefont {Zheng}},\ and\ \bibinfo {author} {\bibfnamefont {Z.}~\bibnamefont {Lu}},\ }\bibfield  {title} {\bibinfo {title} {Stochastic distinguishability of markovian trajectories},\ }\href@noop {} {\bibfield  {journal} {\bibinfo  {journal} {arXiv preprint arXiv:2401.16544}\ } (\bibinfo {year} {2024})}\BibitemShut {NoStop}%
\bibitem [{\citenamefont {Mpemba}\ and\ \citenamefont {Osborne}(1969)}]{mpemba1969cool}%
  \BibitemOpen
  \bibfield  {author} {\bibinfo {author} {\bibfnamefont {E.~B.}\ \bibnamefont {Mpemba}}\ and\ \bibinfo {author} {\bibfnamefont {D.~G.}\ \bibnamefont {Osborne}},\ }\bibfield  {title} {\bibinfo {title} {Cool?},\ }\href@noop {} {\bibfield  {journal} {\bibinfo  {journal} {Physics Education}\ }\textbf {\bibinfo {volume} {4}},\ \bibinfo {pages} {172} (\bibinfo {year} {1969})}\BibitemShut {NoStop}%
\bibitem [{\citenamefont {Lu}\ and\ \citenamefont {Raz}(2017)}]{lu2017nonequilibrium}%
  \BibitemOpen
  \bibfield  {author} {\bibinfo {author} {\bibfnamefont {Z.}~\bibnamefont {Lu}}\ and\ \bibinfo {author} {\bibfnamefont {O.}~\bibnamefont {Raz}},\ }\bibfield  {title} {\bibinfo {title} {Nonequilibrium thermodynamics of the markovian mpemba effect and its inverse},\ }\href@noop {} {\bibfield  {journal} {\bibinfo  {journal} {Proceedings of the National Academy of Sciences}\ }\textbf {\bibinfo {volume} {114}},\ \bibinfo {pages} {5083} (\bibinfo {year} {2017})}\BibitemShut {NoStop}%
\bibitem [{\citenamefont {Klich}\ \emph {et~al.}(2019)\citenamefont {Klich}, \citenamefont {Raz}, \citenamefont {Hirschberg},\ and\ \citenamefont {Vucelja}}]{klich2019mpemba}%
  \BibitemOpen
  \bibfield  {author} {\bibinfo {author} {\bibfnamefont {I.}~\bibnamefont {Klich}}, \bibinfo {author} {\bibfnamefont {O.}~\bibnamefont {Raz}}, \bibinfo {author} {\bibfnamefont {O.}~\bibnamefont {Hirschberg}},\ and\ \bibinfo {author} {\bibfnamefont {M.}~\bibnamefont {Vucelja}},\ }\bibfield  {title} {\bibinfo {title} {Mpemba index and anomalous relaxation},\ }\href@noop {} {\bibfield  {journal} {\bibinfo  {journal} {Physical Review X}\ }\textbf {\bibinfo {volume} {9}},\ \bibinfo {pages} {021060} (\bibinfo {year} {2019})}\BibitemShut {NoStop}%
\bibitem [{\citenamefont {Chittari}\ and\ \citenamefont {Lu}(2023)}]{chittari2023geometric}%
  \BibitemOpen
  \bibfield  {author} {\bibinfo {author} {\bibfnamefont {S.~S.}\ \bibnamefont {Chittari}}\ and\ \bibinfo {author} {\bibfnamefont {Z.}~\bibnamefont {Lu}},\ }\bibfield  {title} {\bibinfo {title} {Geometric approach to nonequilibrium hasty shortcuts},\ }\href@noop {} {\bibfield  {journal} {\bibinfo  {journal} {The Journal of Chemical Physics}\ }\textbf {\bibinfo {volume} {159}} (\bibinfo {year} {2023})}\BibitemShut {NoStop}%
\bibitem [{\citenamefont {McInnes}\ \emph {et~al.}(2018)\citenamefont {McInnes}, \citenamefont {Healy},\ and\ \citenamefont {Melville}}]{mcinnes2018umap}%
  \BibitemOpen
  \bibfield  {author} {\bibinfo {author} {\bibfnamefont {L.}~\bibnamefont {McInnes}}, \bibinfo {author} {\bibfnamefont {J.}~\bibnamefont {Healy}},\ and\ \bibinfo {author} {\bibfnamefont {J.}~\bibnamefont {Melville}},\ }\bibfield  {title} {\bibinfo {title} {Umap: Uniform manifold approximation and projection for dimension reduction},\ }\href@noop {} {\bibfield  {journal} {\bibinfo  {journal} {arXiv preprint arXiv:1802.03426}\ } (\bibinfo {year} {2018})}\BibitemShut {NoStop}%
\end{thebibliography}%
\end{document}